\documentclass[12pt]{article}

\usepackage{graphics,amssymb,float}
\usepackage{graphicx}
\usepackage{caption}
\usepackage{rotating}
\usepackage{dcolumn}
\usepackage{bm}
\usepackage{cite}
\usepackage{amsmath}
\usepackage{indentfirst} 
\usepackage{epstopdf}
\usepackage[usenames,dvipsnames]{xcolor}
\usepackage{setspace}
\usepackage{hhline}
\usepackage{enumitem}

\usepackage{xcolor}
\usepackage{color}
\usepackage{colortbl}

\makeatletter

\@addtoreset{equation}{section}
\makeatother

\def\lsim{\;\raise0.3ex\hbox{$<$\kern-0.75em\raise-1.1ex\hbox{$\sim$}}\;}
\def\gsim{\;\raise0.3ex\hbox{$>$\kern-0.75em\raise-1.1ex\hbox{$\sim$}}\;}
\def\beq{\begin{equation}}   \def\eeq{\end{equation}}
\def\ba{\begin{array}}       \def\ea{\end{array}}
\def\bea{\begin{eqnarray}}   \def\eea{\end{eqnarray}}

\def\nl{\newline}

\begin{document}

\begin{titlepage}


\begin{center}
\vspace{1cm}

{\Large\bf NMSSM with correct relic density and an additional 95~GeV Higgs boson}

\vspace{2cm}

{\bf{Ulrich Ellwanger$^a$ and Cyril Hugonie$^b$}}\\
\vspace{1cm}
\it $^a$ IJCLab,  CNRS/IN2P3, University  Paris-Saclay, 91405  Orsay,  France\\
ulrich.ellwanger@ijclab.in2p3.fr\\
\it $^b$ LUPM, UMR 5299, CNRS/IN2P3, Universit\'e de Montpellier, 34095 Montpellier, France\\
cyril.hugonie@umontpellier.fr

\end{center}
\vspace{2cm}

\begin{abstract}

We investigate whether it is possible within the NMSSM to describe simultaneously a correct dark matter relic density complying with the latest null results from the LZ experiment, an extra Higgs boson with a mass of $\sim 95$~GeV visible in the bb channel at LEP and the diphoton channel at the LHC, and the deviation of the anomalous magnetic moment of the muon from its value within the Standard Model. We find that this is still possible for a variety of dark matter annihilation mechanisms, for singlino-like but also bino-like lightest supersymmetric particles. We show the signal rates of the extra Higgs boson and the dark matter detection rates as function of the dark matter mass and annihilation mechanism. The dark matter direct detection cross section may well fall below the neutrino floor. The masses of the lightest electroweakly interacting supersymmetric particles are typically not far above 100~GeV, but not excluded due to unconventional decays.

\end{abstract}

\end{titlepage}

\section{Introduction}

Supersymmetric extensions of the Standard Model may explain the dark matter of the universe. The lightest supersymmetric partner particle is stable, provided R-parity is conserved. It can well be neutral and freeze out into the desired relic density $\Omega_{DM} h^2 = 0.1187$ (we add $\pm 10\%$ theoretical uncertainty) in agreement with WMAP/Planck results \cite{Hinshaw:2012aka,Ade:2013zuv}. However, it also has to comply with the absence of signals in direct detection experiments in spin-independent channels by the CRESST \cite{CRESST:2015txj}, PandaX-II \cite{Tan:2016zwf}, LUX \cite{Akerib:2016vxi}, XENON1T \cite{Aprile:2017iyp,Aprile:2018dbl}, DarkSide \cite{DarkSide:2018bpj} and LZ \cite{LZ:2022lsv} collaborations, and in spin-dependent channels by the PICO-2L \cite{Amole:2016pye}, LUX \cite{Akerib:2016lao} and PandaX-II \cite{Fu:2016ega}, Xenon1T \cite{Aprile:2017iyp,Aprile:2018dbl} and LZ \cite{LZ:2022lsv} collaborations.

Within the Minimal Supersymmetric extension of the Standard Model (MSSM) it is no longer easy to comply with these constraints \cite{Baer:2016ucr}; in \cite{Baer:2018rhs} it has been argued that additional contributions to the dark matter relic density would be required, but co-annihilation of higgsinos with charginos, with masses beyond 1~TeV, may still be a possibility \cite{Delgado:2020url}.
In \cite{Chakraborti:2021mbr} it has been observed that co-annihilation with charginos or sleptons can allow for dark matter made out of a Lightest Supersymmetric Particle (LSP) consisting of a bino/wino-mixture, however at the price of some fine tuning \cite{Wang:2022rfd,Gomez:2022qrb,Yang:2022gvz} (see also \cite{Chakraborti:2021kkr,Iwamoto:2021aaf,VanBeekveld:2021tgn,Cox:2021nbo,Chakraborti:2021bmv,Baer:2021aax,Shafi:2021jcg,Forster:2021vyz,Agashe:2022uih,Zhao:2022pnv,Baum:2023inl,He:2023lgi,Bisal:2023iip}). Staus degenerate with the LSP seem also possible \cite{Chakraborti:2023pis}. Given the lower bounds on chargino and slepton masses already from LEP, the mass of the LSP has then to be above 100~GeV. In \cite{Barman:2024xlc} it is shown that, considering the constraints from LZ \cite{LZ:2022lsv} and the LHC, even if additional contributions to the dark matter relic density are allowed the MSSM requires heavy higgsinos, or narrow regions in parameter space with light higgsinos or light staus.

Such constraints are alleviated within the Next-to-Minimal extension of the Standard Model (NMSSM) \cite{Ellwanger:2009dp} where the LSP can have at least a strong singlino-component and correspond to the fermionic superpartner of the gauge singlet superfield. A pure singlino has no couplings to quarks and leptons, hence such an LSP has very small direct detection cross sections which can fall below the projections for the exclusion reach of the XENON-nT \cite{XENON:2020kmp} and LZ \cite{LZ:2018qzl}, or even below the neutrino floor \cite{Billard:2013qya}.
Hence the singlino is a quite attractive dark matter candidate \cite{Cerdeno:2004xw,Belanger:2005kh,Cerdeno:2007sn,Barger:2007nv,Belanger:2008nt,Vasquez:2010ru,Perelstein:2012qg,Kozaczuk:2013spa,Cao:2013mqa,Kim:2014noa,Ellwanger:2014dfa,Ishikawa:2014owa,Han:2014nba,Cheung:2014lqa,Huang:2014cla,Cahill-Rowley:2014ora,Guo:2014gra,Cao:2014efa,Bi:2015qva,Cao:2015loa,Butter:2015fqa,Gherghetta:2015ysa,Han:2015zba,Potter:2015wsa,Barducci:2015zna,Enberg:2015qwa,Badziak:2015exr,Xiang:2016ndq,Cao:2016nix,Cao:2016cnv,Beskidt:2017xsd,Badziak:2017uto,Mou:2017sjf,
Ellwanger:2016sur,Baum:2017enm,Shang:2018dja,Ellwanger:2018zxt,Cao:2018rix,Domingo:2018ykx,Cao:2019qng,Abdallah:2019znp,Wang:2019biy,Wang:2020tap,Wang:2020dtb,Guchait:2020wqn,Barman:2020vzm,Wang:2020xta,Zhou:2021pit,Cao:2021ljw,Cao:2021tuh,Cao:2022chy,Chatterjee:2022pxf,Cao:2022htd,Almarashi:2022iol,Cao:2022ovk,Wang:2023suf,Heng:2023xfb,Cao:2023juc,Cao:2024axg} which can remain consistent with the absence of signals in direct detection experiments.

Actually hints exist for additional Higgs bosons as present in the NMSSM. The combination of searches for the SM Higgs boson by the ALEPH, DELPHI, L3 and OPAL experiments at LEP \cite{LEPWorkingGroupforHiggsbosonsearches:2003ing} showed some mild excess of events in the $Z^*\to Z+b\bar{b}$ channel in the mass region $95-100$~GeV.  
Searches for Beyond-the-Standard Model (BSM) Higgs bosons at the LHC in the diphoton channel were performed by CMS and ATLAS. A search at run~1 by CMS showed a $\sim 2\, \sigma$ excess at 97~GeV \cite{CMS-PAS-HIG-14-037}, which was confirmed by CMS later in \cite{CMS:2018cyk} and in \cite{CMS-PAS-HIG-20-002} for a mass hypothesis of 95.4~GeV. A somewhat less sensitive search in the diphoton channel by ATLAS in \cite{ATLAS-CONF-2018-025} lead to an upper limit on the fiducial cross section which did not contradict the possible excess observed by CMS, a recent analysis by ATLAS in the diphoton channel in \cite{ATLAS-CONF-2023-035} showed a mild excess of $1.7\, \sigma$ at 95~GeV.
A search for BSM Higgs bosons in the di-tau channel by CMS in \cite{CMS:2022goy} showed an excess of $2.6\, \sigma$ (local) for a mass of $ 95-100$~GeV.

The $\sim 2\, \sigma$ excess at LEP was quantified in \cite{Cao:2016uwt}. Let us denote the extra (lighter) Higgs boson by $H_{95}$, with a reduced coupling to vector bosons $W^\pm, Z$ (relative to the coupling of a SM-like Higgs boson of corresponding mass) given by $C_V(H_{95})$. Then the authors in \cite{Cao:2016uwt} define
\beq\label{mulep}
\mu^{LEP}_{bb} \equiv C_V(H_{95})^2\times BR(H_{95}\to b\bar{b})/BR(H_{SM}^{95}\to b\bar{b})= 0.117 \pm 0.057
\eeq
where $H_{SM}^{95}$ denotes a fictitious SM-like Higgs boson with a mass of 95~GeV.

The best fits for a diphoton signal of $H_{95}$ in CMS and ATLAS were combined in \cite{Biekotter:2023oen}. The authors in \cite{Biekotter:2023oen} obtain
\beq\label{mugamgam}
\mu_{\gamma\gamma}^{LHC} = \frac{\sigma(gg \to H_{95}\to \gamma\gamma)}{\sigma(gg \to H_{SM}^{95} \to \gamma\gamma)}
=  {0.24^{+0.09}_{-0.08}}\; .
\eeq
Again, $H_{SM}^{95}$ denotes a SM-like Higgs boson with a mass of $\sim 95$~GeV.

The best fit for the excess in the di-tau channel at 95~GeV observed by CMS in \cite{CMS:2022goy} corresponds to a cross section times branching fraction
\beq\label{ditau}
\sigma(gg \to H_{95} \to \tau\tau)= 7.8^{+3.9}_{-3.1}\text{pb}\; .   
\eeq
Relative to $H_{SM}^{95}$ we obtain
\beq\label{ditau1}
\mu_{\tau\tau}^{LHC} = \frac{\sigma(gg \to H_{95}\to \tau\tau)}{\sigma(gg \to H_{SM}^{95} \to \tau\tau)}
= 1.38^{+0.69}_{-0.55}\; .
\eeq

Such an additional Higgs boson with a mass of $\sim 95$~GeV has been discussed within the NMSSM in
\cite{Belanger:2012tt,Cao:2016uwt,Biekotter:2017xmf,Hollik:2018yek,Domingo:2018uim,Wang:2018vxp,Cao:2019ofo,Choi:2019yrv,Hollik:2020plc,Biekotter:2021qbc,Adhikary:2022pni,Li:2022etb,Biekotter:2023oen,Ellwanger:2023zjc,Cao:2023gkc,Li:2023kbf,Roy:2024yoh}; mostly, however, allowing for a too small dark matter relic density (assuming additional contributions), and/or before the consideration of constraints from LZ \cite{LZ:2022lsv}. It is the aim of the present paper to verify whether attractive scenarios within the NMSSM exist with a correct relic density, satisfying the LZ-constraints, and an additional Higgs boson with a mass of $\sim 95$~GeV. To this end we scan the parameter space over the NMSSM-parameters $\lambda$, $\kappa$, $A_\lambda$, $A_\kappa$, $\mu_{\text{eff}}$, $\tan\beta$ 
in the Higgs sector and non universal soft supersymmetry breaking terms in the gaugino/sfermion sector; for their definitions we refer to review in \cite{Ellwanger:2009dp}.
We employ the codes \texttt{NMSSMTools-6.0.2} \cite{Ellwanger:2004xm,Ellwanger:2005dv,NMSSMTools} and {\sf micrOMEGAs$\_$3}~\cite{Belanger:2013oya}.

We impose constraints from b-physics and constraints from searches for BSM Higgs bosons by ATLAS and CMS as implemented in \texttt{NMSSMTools-6.0.2}. The references to constraints from BSM Higgs-boson searches and b-physics (of little relevance here) are listed on the web page {\sf https://www.lupm.in2p3.fr/users/nmssm/history.html}. Constraints from the absence of a Landau singularity for the Yukawa couplings below the GUT scale confine values of the NMSSM-specific coupling $\lambda$ to $\lambda \lsim 0.7$ (see \cite{Ellwanger:2009dp} for the the meaning of NMSSM specific parameters). 

Constraints from the anomalous magnetic moment of the muon $a_\mu$ \cite{Muong-2:2006rrc,Muong-2:2021ojo} as in \cite{Domingo:2022pde} can always be satisfied by choosing the soft supersymmetry breaking smuon, bino and/or wino masses small enough, and the trilinear coupling $A_\mu$ large enough. These parameters have little impact on the quantities of relevance here which are shown in the Figures in the next Section, hence constraints from $a_\mu$ are not imposed on most of the points in parameter space. However, for the benchmark points in the next Section, constraints from the $a_\mu$ are satisfied.
The constraint on $M_W$ as applied in \cite{Domingo:2022pde} is not used since it relies on a single experimental result which differs significantly from many others. 
All soft supersymmetry breaking terms are taken below 5~TeV.

Constraints on the sparticle spectrum from the absence of signals at the LHC are taken into account using the code \texttt{SModels-2.2.0} \cite{Kraml:2013mwa,Dutta:2018ioj,Khosa:2020zar,Alguero:2021dig}. Constraints on light electroweakly interacting sparticles (in the $\sim 100 - 200$~GeV range) from the LHC rely on production cross sections for higgsinos or winos, and their decays into W- or Z-bosons, see e.g. the recent ATLAS combination in \cite{ATLAS:2024qxh}. In the NMSSM, both production cross sections of light electroweakly interacting sparticles as well as corresponding decay rates can be suppressed and the constraints from the LHC become weak. Moreover, in the presence of several light electroweakly interacting sparticles, heavy sparticles (squarks, gluino) can undergo several distinct cascade decays such that the probability for each of them is reduced. As a consequence such scenarios are difficult to rule out.

We require that the singlet-like scalar has a mass in the range $95.4\pm 3$~GeV (allowing for a theoretical uncertainty of 3~GeV), $\mu^{LEP}_{bb}$ in the $2\, \sigma$ range of \eqref{mulep}, and $\mu_{\gamma\gamma}^{LHC}$ in the $2\, \sigma$ range of \eqref{mugamgam}. 
For the SM-like Higgs boson we require a mass within $125.2\pm 3$~GeV (allowing for theoretical uncertainties), and that the couplings in the $\kappa$-framework satisfy combined limits of CMS \cite{CMS:2022dwd} and ATLAS \cite{ATLAS:2022vkf}. These limits require that the reduced couplings $C_V(H_{SM}$) of the SM-like Higgs boson to gauge fields satisfy $C_V(H_{SM}$)~$\gsim$~0.96, hence the mixing angle between $H_{95}$ and $H_{SM}$ is limited such that the reduced coupling of $H_{95}$ to $\tau\tau$ is bounded to below $25\%$. Then the (reduced) cross section $\times$ BR($H_{95}\to \tau\tau$) can hardly exceed 0.1 in contrast to what is desired within the $2\, \sigma$ range for the di-tau excess $\mu_{\tau\tau}^{LHC}$ in \eqref{ditau1}; its description within the $2\, \sigma$ range is therefore left aside since impossible for the type II Yukawa couplings present in the (N)MSSM.

In the next Section we present our results in the form of Figures and benchmark points, Section~3 is devoted to conclusions.


\section{Results}

Various mechanisms exist within the NMSSM for a sufficiently fast annihilation of dark matter in the form of a singlino-like LSP:

\begin{enumerate}[label=(\alph*)]

\item Annihilation via a pseudoscalar in the s-channel. Then $M_{A_1}$ should be about $2\times M_{\chi_1^0}$ such that the annihilation cross section is enhanced by the s-channel pole. We find that this annihilation mechanism is possible in the NMSSM for $M_{\chi_1^0}\lsim 24$~GeV, i.e. $M_{A_1}\lsim 50$~GeV. $A_1$ is always mostly singlet-like, and decays into $b\bar{b}$ for $M_{A_1} \gsim 9$~GeV, into $\tau\tau$ otherwise. 

\item Annihilation via the Standard~Model-like Higgs boson $H_{SM}$ or the mostly singlet-like Higgs boson $H_{95}$ in the s-channel if the singlino mass is about half the corresponding mass. (With $M_{H_{95}}\sim 95$~GeV, the $Z$-boson plays little role in the s-channel since the couplings of an LSP with $M_{LSP}\sim M_Z/2$ to $H_{95}$ are typically much stronger.)

\item Co-annihilation with charginos, sleptons, or the next-to-lightest neutralinos. (Subsequently we include staus and their sneutrinos in the notion sleptons.) Often several of these sparticles intervene simultaneously with different strengths; then these sparticles must be close in mass (within $\sim 10$~GeV) to the singlino. In most cases, the lightest charged or electroweakly interacting sparticle must be heavier than $\sim 100$~GeV from constraints from LEP. Then a lower bound of $\sim 90$~GeV holds for the singlino, and the properties of the valid points shown in the figures below depend little on the nature of the sparticle(s) which intervene in co-annihilation. An exception are sneutrinos which can be lighter, as light as 55~GeV if they annihilate through the $H_{SM}$ funnel.

\item A bino-like LSP as in the MSSM under the latest constraints \cite{Barman:2024xlc} (or in the NMSSM e.g. in \cite{Abdallah:2020yag,Datta:2022bvg}). For a practically pure bino, LEP and LHC constraints allow for masses up to 330~GeV down to $\sim 20$~GeV. Compared to singlino-like LSPs, pair-annihilation of binos via a (pseudo-)scalar in the s-channel is relatively less important. The reason is that singlino-singlino-Higgs Yukawa couplings $\sim \lambda$ exist in the NMSSM for the singlet-like (pseudo-)scalar at tree level, and other scalars through mixing. However, bino-bino-Higgs Yukawa couplings do not exist for pure binos since the bino is a gaugino of an abelian gauge symmetry. The dominant annihilation channelis therefore co-annihilation with sleptons.

\end{enumerate}

All mechanisms allow for a dark matter relic density in agreement with the WMAP/Planck result $\Omega_{DM} h^2 = 0.1187 \pm 10\%$.

It is interesting to compare the possible values of the signal strengths $\mu^{LEP}_{bb}$, $\mu_{\gamma\gamma}^{LHC}$ and $\mu_{\tau\tau}^{LHC}$ as well as the dark matter cross sections (spin independent on protons $\sigma_{p}^{SI}$, and spin dependent on neutrons $\sigma_{n}^{SD}$) for the various possible dark matter masses associated to various annihilation processes, as function of the mass of the LSP.

In Figs.~1 we show in how far points in the NMSSM can accomodate an additional Higgs boson near 95~GeV with signal rates $\mu^{LEP}_{bb}$ at LEP as in eq.~\eqref{mulep}. (Only points within the $2\,\sigma$ range of \eqref{mulep} are shown.) 
Here and in the subsequent figures, points on the left hand side refer to a singlino-like LSP. Points corresponding to annihilation via a pseudoscalar $A_1$ (A1 funnel) are shown in violet, annihilation via $H_{95}$ or $H_{SM}$ in orange, co-annihilation with charginos, sleptons or next-to-lightest neutralinos in blue. 
Note that, for a LSP mass near $H_{SM}/2$ or $H_{95}/2$, the dominant dark matter annihilation mechanism can be either via the corresponding Higgs funnel or via co-annihilation.
Points with a bino-like LSP are shown on the right hand side in black. The red dots denote benchmark points whose details are given in the Table~2-4 below. Comparing the right hand side and the left hand side of Figs.~1 one observes that the possible values of $\mu^{LEP}_{bb}$ are generally somewhat larger for bino LSPs compared to singlino LSPs.

\begin{figure}[ht!]
\begin{center}
\hspace*{-5mm}
\begin{tabular}{cc}
\includegraphics[scale=0.3]{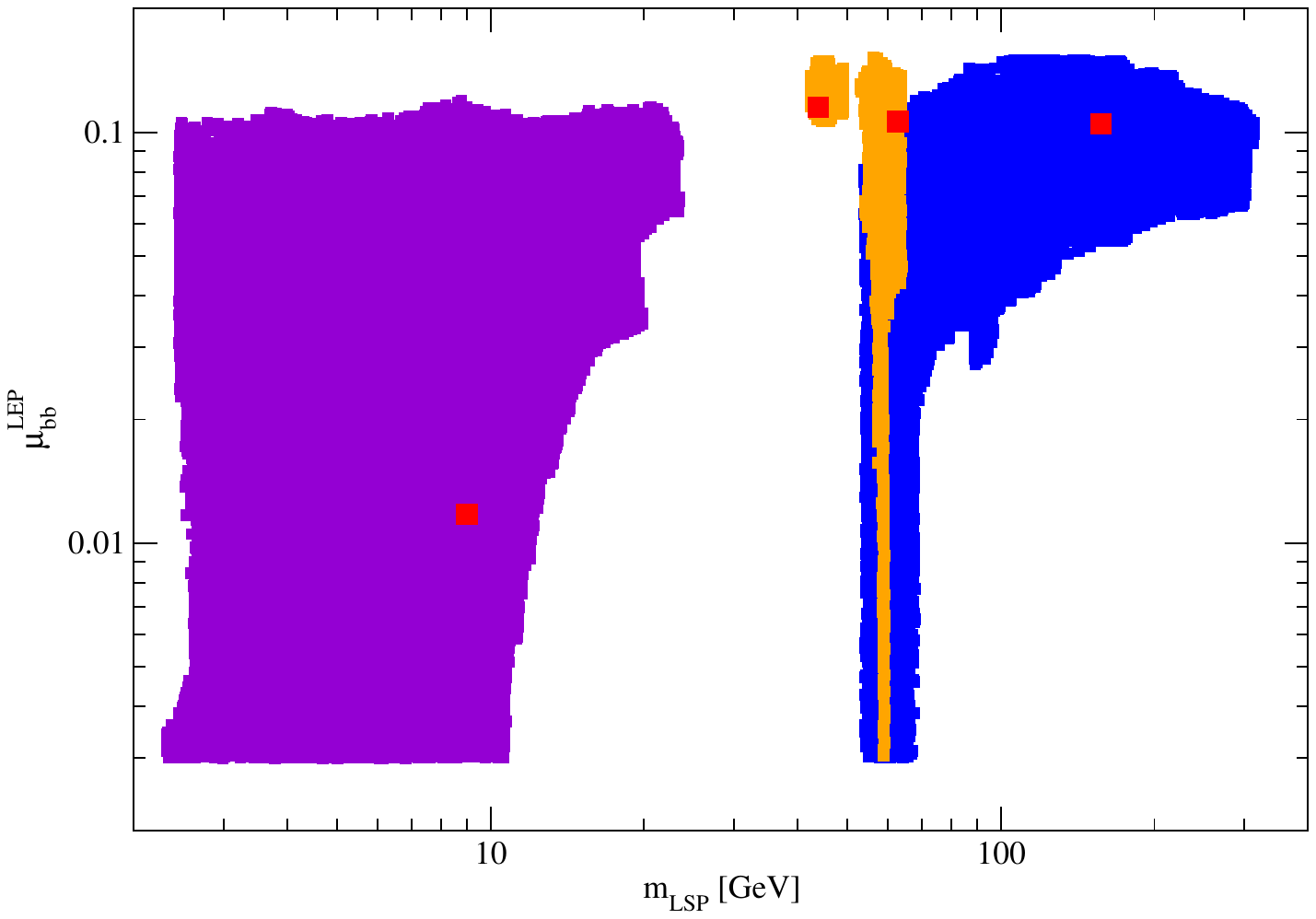}
   & 
\includegraphics[scale=0.3]{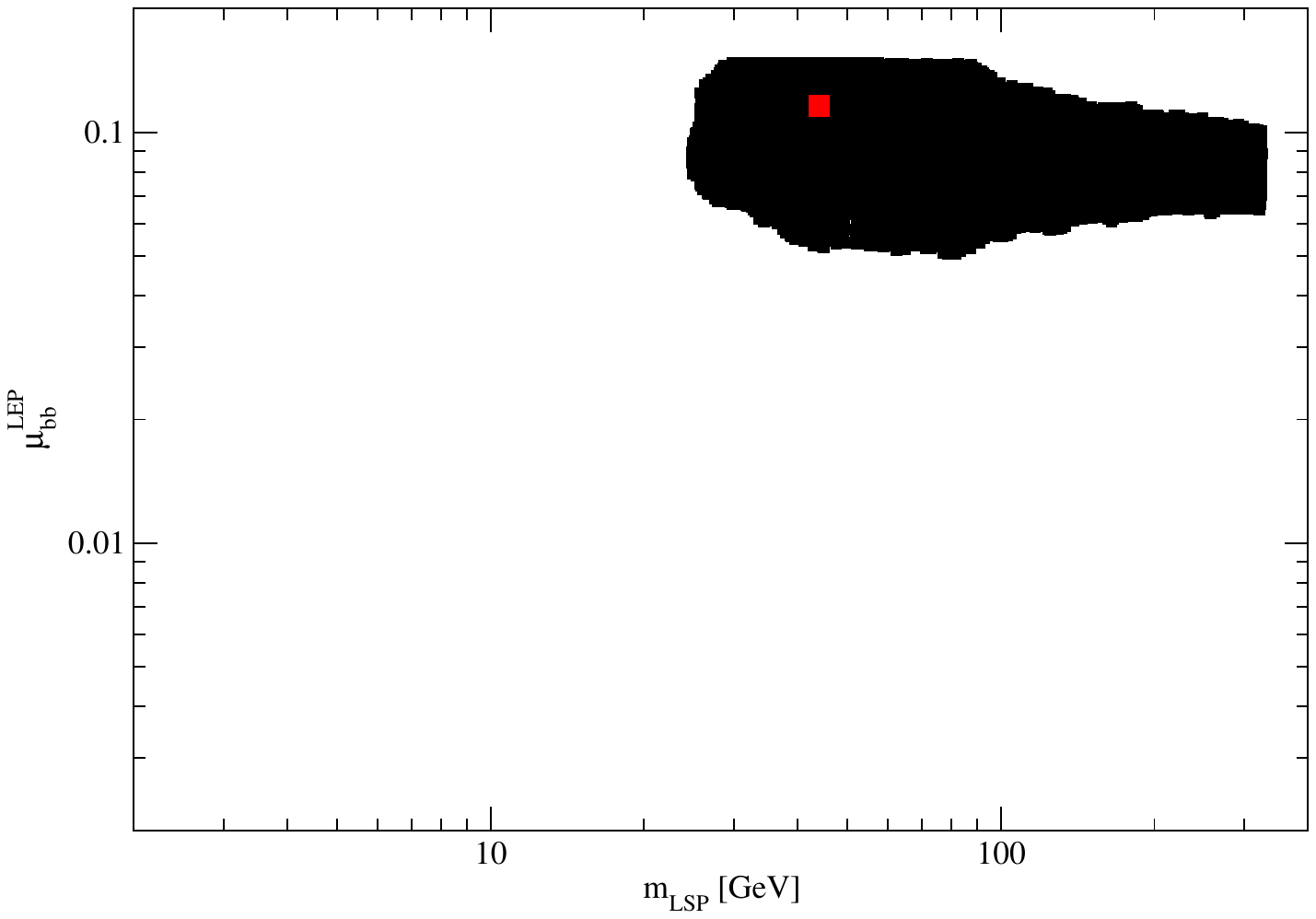}
\end{tabular}
\end{center}
\caption{$\mu^{LEP}_{bb}$  for points in the NMSSM satisfying all constraints. 
Left hand side: Points with a singlino-like LSP. Points corresponding to annihilation via a pseudoscalar $A_1$ (A1 funnel) are shown in violet, annihilation via $H_{95}$ or $H_{SM}$  in orange, co-annihilation with charginos, sleptons, next-to-lightest neutralinos in blue.
Right hand side: Points with a bino-like LSP. The red dots denote benchmark points whose details are given in the Tables~2-4 below.}
\label{fig:1}
\end{figure}

The contributing range of parameters depends somewhat on the annihilation mechanism. In Table~1 we give these ranges for the NMSSM specific parameters and the soft supersymmetry breaking masses $M_1$, $M_2$ and $M_3$ where the latter refer to the bino, wino and gluino masses, respectively. The soft supersymmetry breaking mass terms for scalars vary from 83~GeV (for sleptons) and 320~GeV (for squarks) up to 5~TeV, the soft supersymmetry breaking trilinear couplings from $-5$~TeV to 5~TeV. (The upper limits of 5~TeV are imposed by hand.) Given the possibly complicated squark decay cascades, squarks can be relatively light without contradicting constraints from searches at the LHC. In any case, due to radiative corrections the pole squark masses can differ considerably from the soft supersymmetry breaking mass terms.

\begin{table} [ht!]
\begin{center}
\begin{tabular}{| c | c | c | c | c |}
\hline
  & $A_1$-funnel  & $H_{95}$- or $H_{SM}$-funnels & co-annihilation & bino LSP    \\
\hline
$\lambda$       & 0.17$-$0.52&0.18$-$0.32&0.019$-$0.34&0.019$-$0.22 \\
\hline
$\kappa$        &-(6.3$-$0.36)$\times 10^{-3}$ &-(1.48$-$0.59)$\times 10^{-2}$&-0.011$-$0.024& 0.0019$-$0.020\\
\hline
$A_\lambda$     &2862$-$5000&3486$-$5000&-5000$-$4320& -(5000$-$1503) \\
\hline
$A_\kappa$      &12$-$65 &-122$-$175&-3.4$-$551& -1.0$-$585 \\
\hline
$\mu_{\rm eff}$ &350$-$1388&575$-$946&-1326$-$136& -(1226$-$361) \\
\hline
$\tan\beta$     &3.3$-$11&4.0$-$7.7&3.2$-$14& 3.8$-$19   \\
\hline
$M_1$           &15$-$5000&-227$-$5000&56$-$5000 & 25$-$329 \\
\hline
$M_2$           &89$-$5000& 91$-$5000&-5000$-$680& -(5000$-$89) \\
\hline
$M_3$           &746$-$5000&729$-$5000&722$-$5000& 721$-$5000 \\
\hline
\end{tabular}
\caption{Range of NMSSM specific input parameters and bino ($M_1$), wino ($M_2$) and gluino ($M_3$) masses for the various dark matter annihilation mechanisms. All dimensionful parameters are given in GeV.}
\end{center}
\label{tab:1}
\end{table}

In Fig.~2 we show $\mu_{\gamma\gamma}^{LHC}$ for points within the $2\,\sigma$ range of the signal rate \eqref{mugamgam} for an additional Higgs boson near 95~GeV in the $\gamma\gamma$ channel at the LHC. The colors denote the dark matter annihilation mechanisms as in Fig.~1.

\begin{figure}[ht!]
\begin{center}
\hspace*{-5mm}
\begin{tabular}{cc}
\includegraphics[scale=0.3]{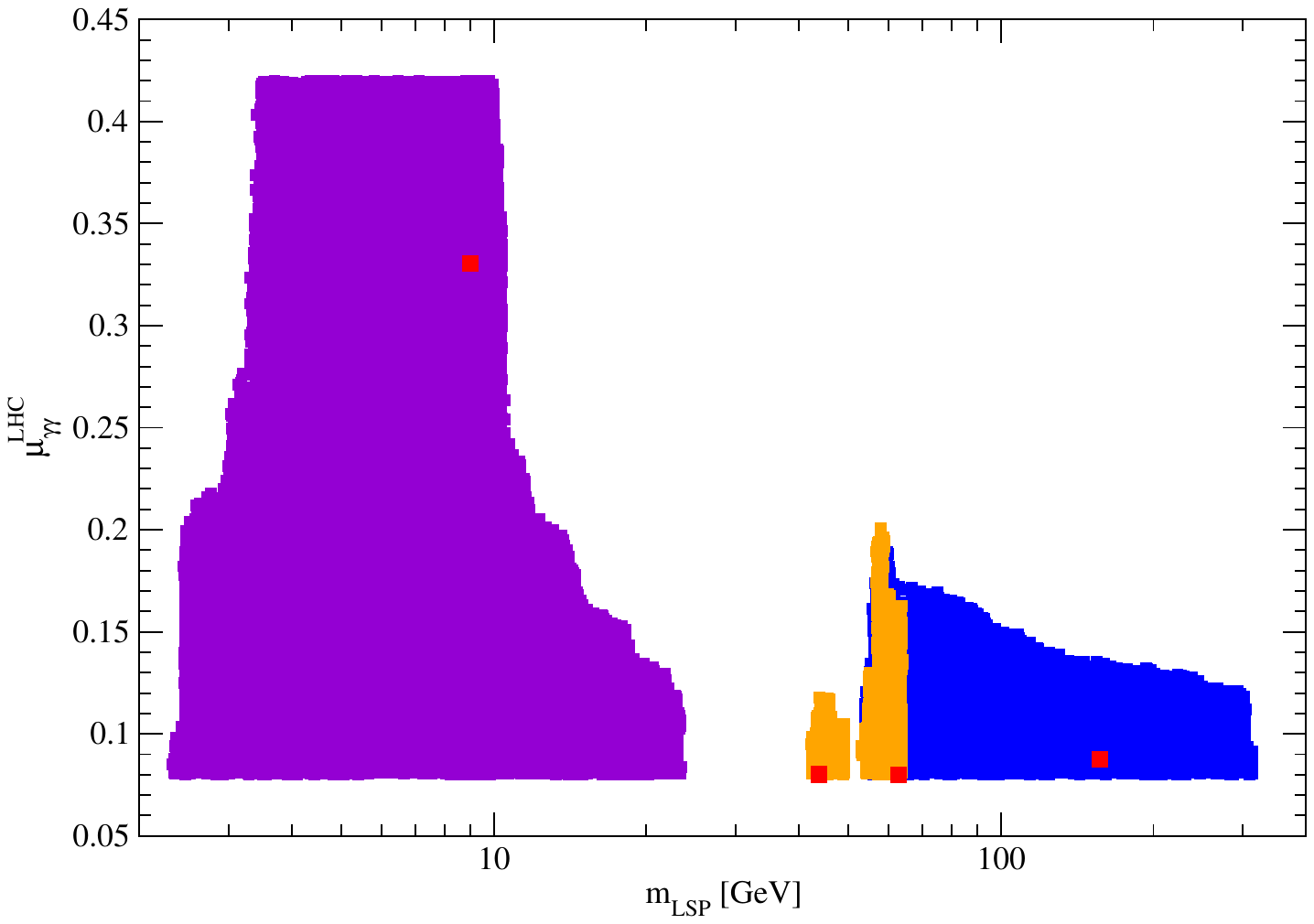}
   & 
\includegraphics[scale=0.3]{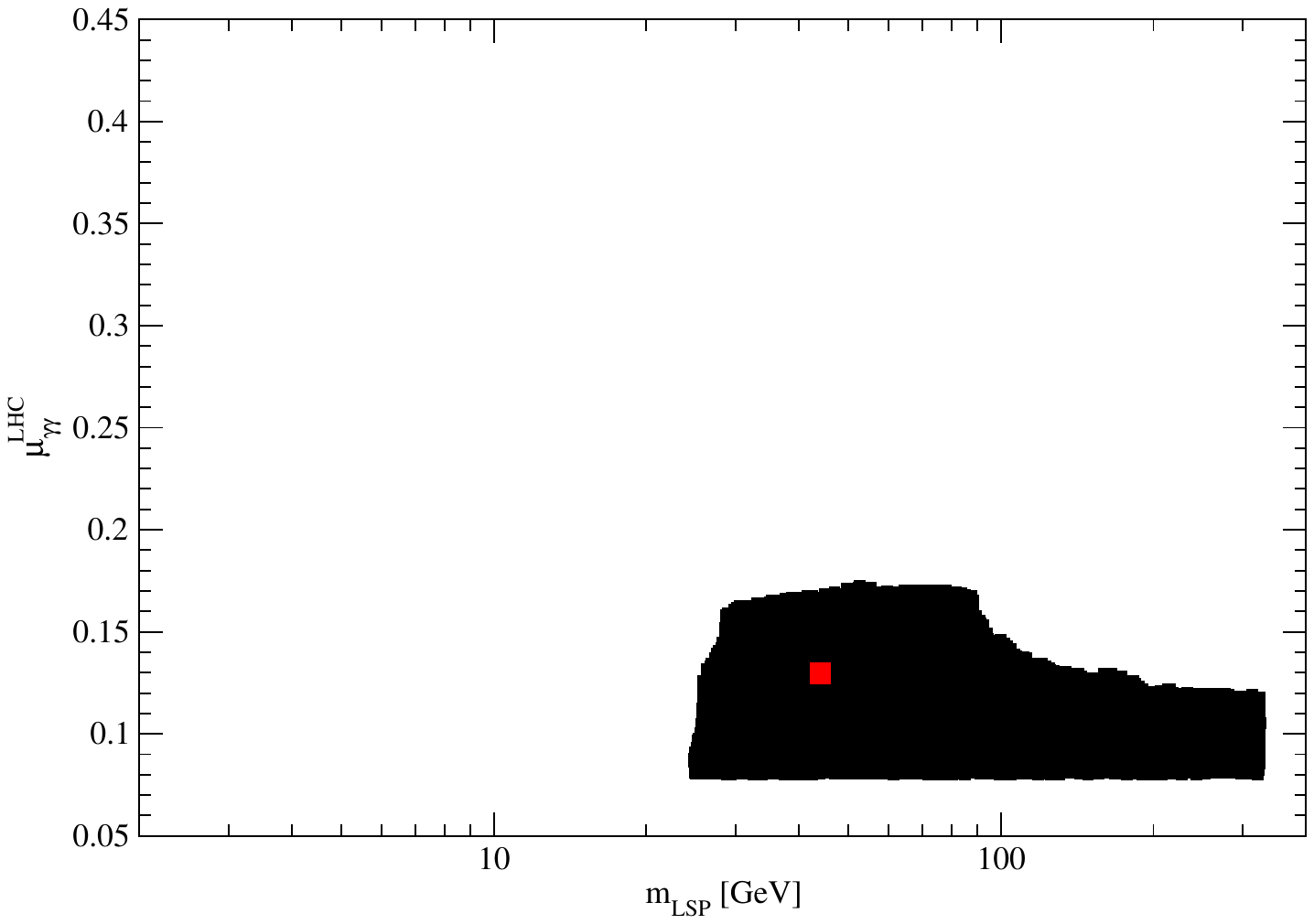}
\end{tabular}
\end{center}
\caption{$\mu^{LHC}_{\gamma\gamma}$  for an additional Higgs boson near 95~GeV for points in the NMSSM satisfying all constraints. The colors denote the dark matter annihilation mechanisms as in Fig.~1.}
\label{fig:2}
\end{figure}

\begin{figure}[ht!]
\begin{center}
\hspace*{-5mm}
\begin{tabular}{cc}
\includegraphics[scale=0.3]{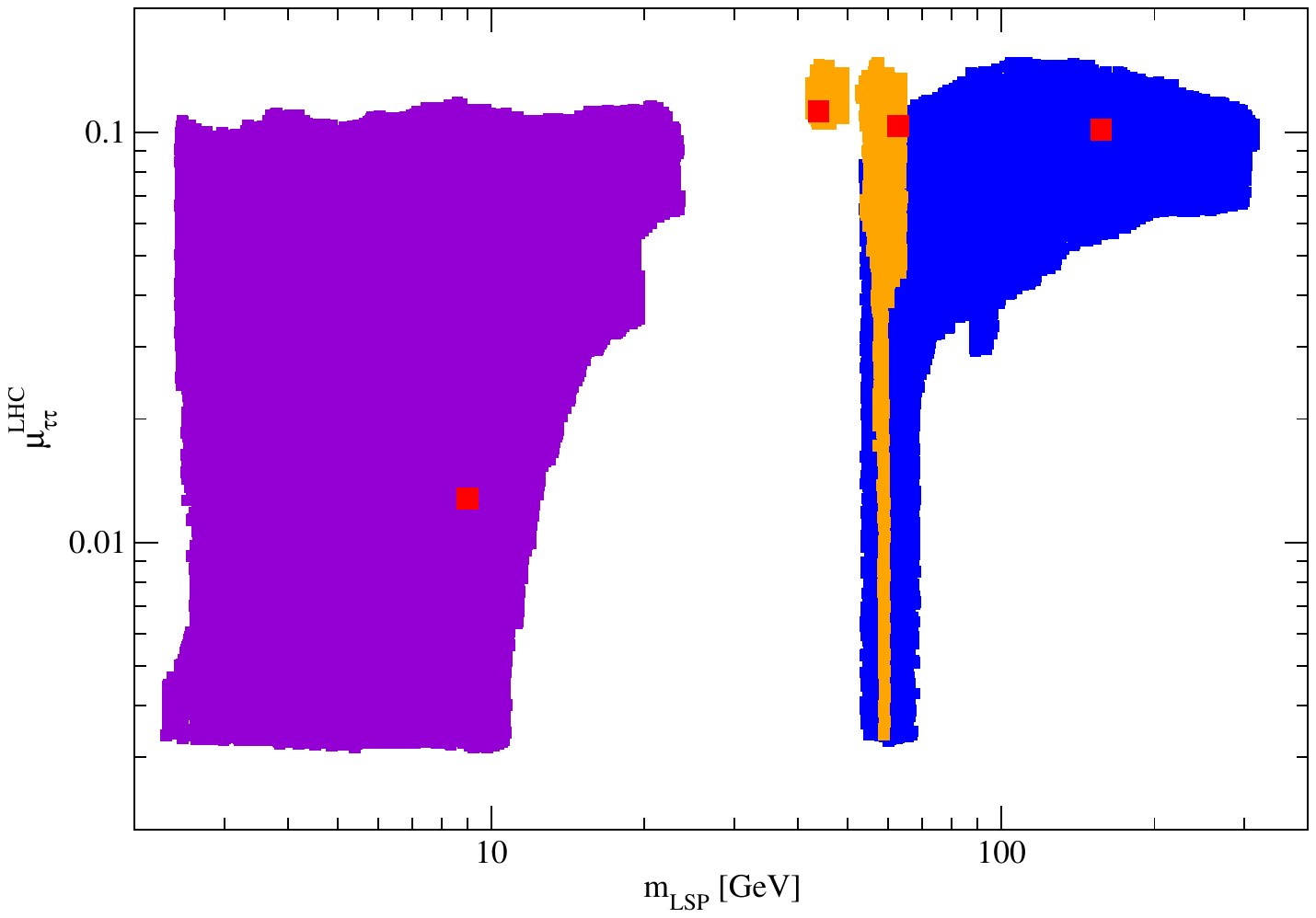}
   & 
\includegraphics[scale=0.3]{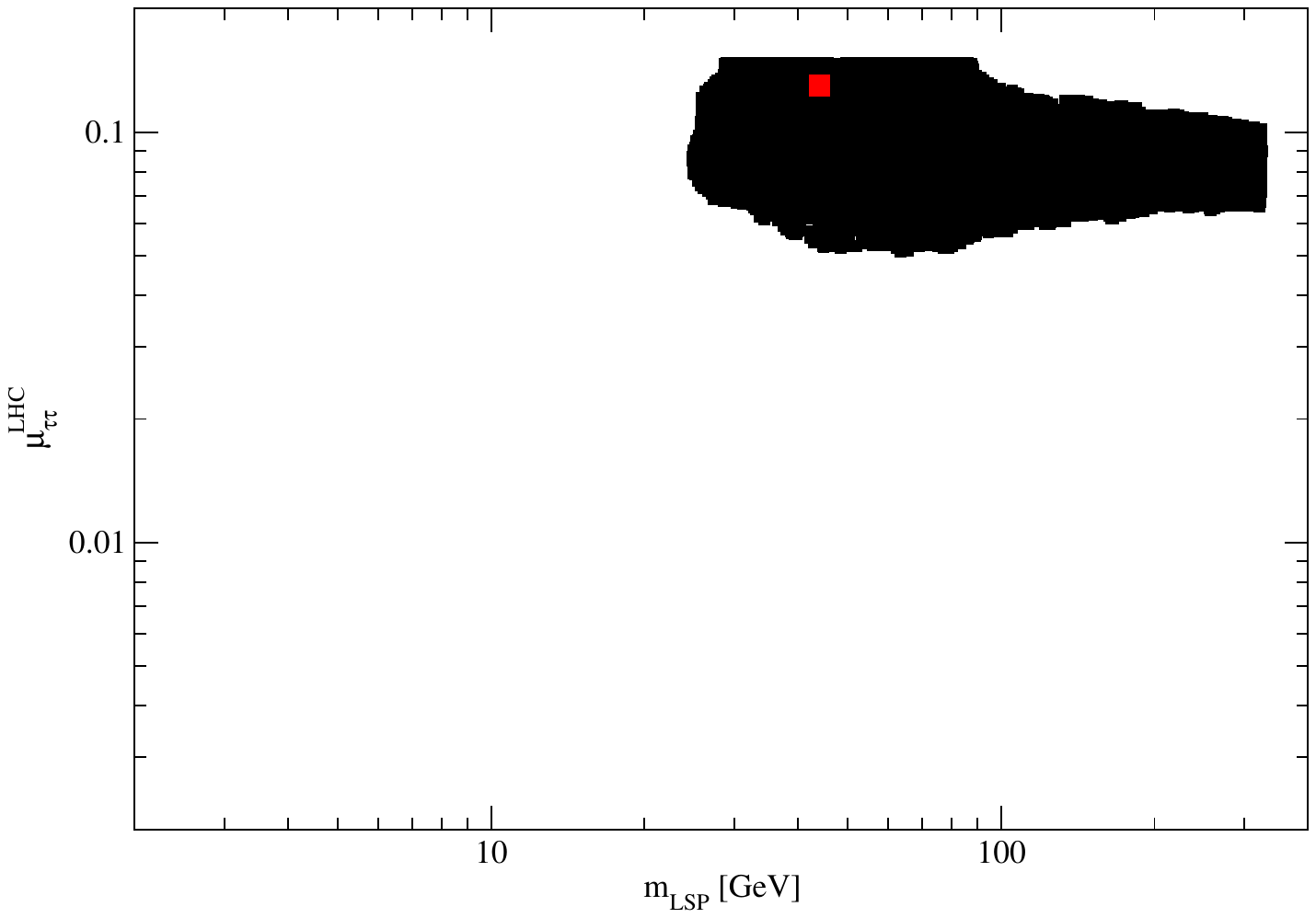}
\end{tabular}
\end{center}
\caption{$\mu^{LHC}_{\tau\tau}$  for an additional Higgs boson near 95~GeV for points in the NMSSM satisfying all constraints. The colors denote the dark matter annihilation mechanisms as in Fig.~1.}
\label{fig:3}
\end{figure}

Fig.~3 shows the signal rate $\mu^{LHC}_{\tau\tau}$ for an additional Higgs boson near 95~GeV for points in the NMSSM within the $2\,\sigma$ ranges of the signal rates \eqref{mulep} and \eqref{mugamgam}, and all other constraints. Actually Fig.~3 looks similar to Fig.~1, the reason being that the couplings of $H_{95}$ to b-quarks and $\tau$-leptons are related in the NMSSM with its type~2 Yukawa couplings of $H_{SM}$; the Yukawa couplings of $H_{95}$ are inherited from those of $H_{SM}$ through mixing. A larger mixing angle would enhance the couplings of $H_{95}$, but reduce simultaneously those of $H_{SM}$. The latter are limited from below by the latest measurements of ATLAS \cite{ATLAS:2022vkf} and CMS \cite{CMS:2022dwd}. Hence it is difficult in the NMSSM to obtain large couplings of $H_{95}$, i.e. larger signal rates than those shown in Fig.~3 which are below the $2\,\sigma$ range of 0.28 for $\mu^{LHC}_{\tau\tau}$ in \eqref{ditau1}.

In Fig.~4 we show the spin-independent (SI) and spin-dependent (SD) direct detection cross section. We also show the present exclusion of the LZ experiment \cite{LZ:2022lsv}, projections for the LZ exclusion reach from \cite{LZ:2018qzl}, and the neutrino floor from \cite{Billard:2013qya}. We see that the SI direct detection cross section can well fall below the neutrino floor, in which case direct detection experiments are not able to verify this form of dark matter. 

\begin{figure}[ht!]
\begin{center}
\hspace*{-5mm}
\begin{tabular}{cc}
\includegraphics[scale=0.3]{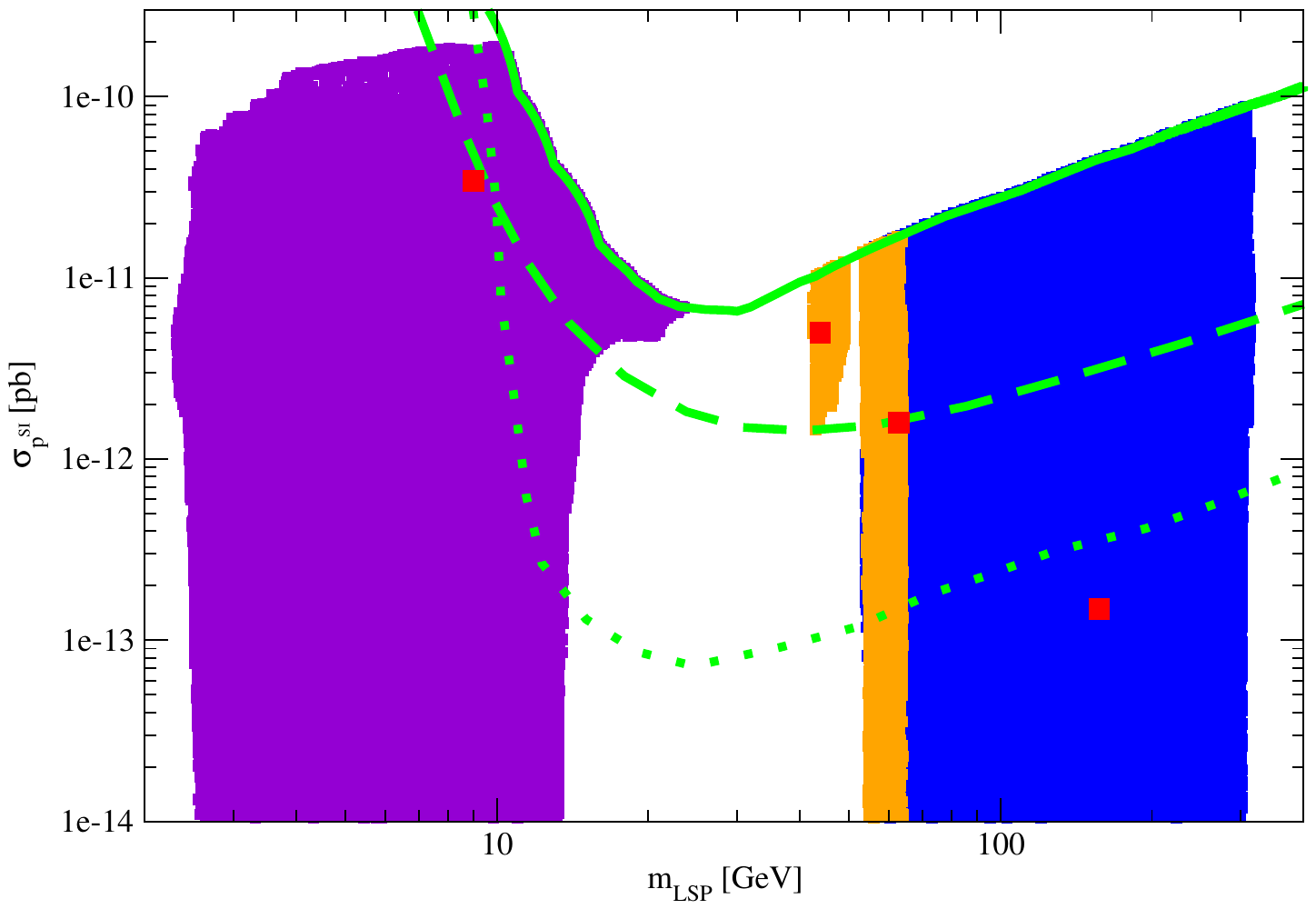}
   & 
\includegraphics[scale=0.3]{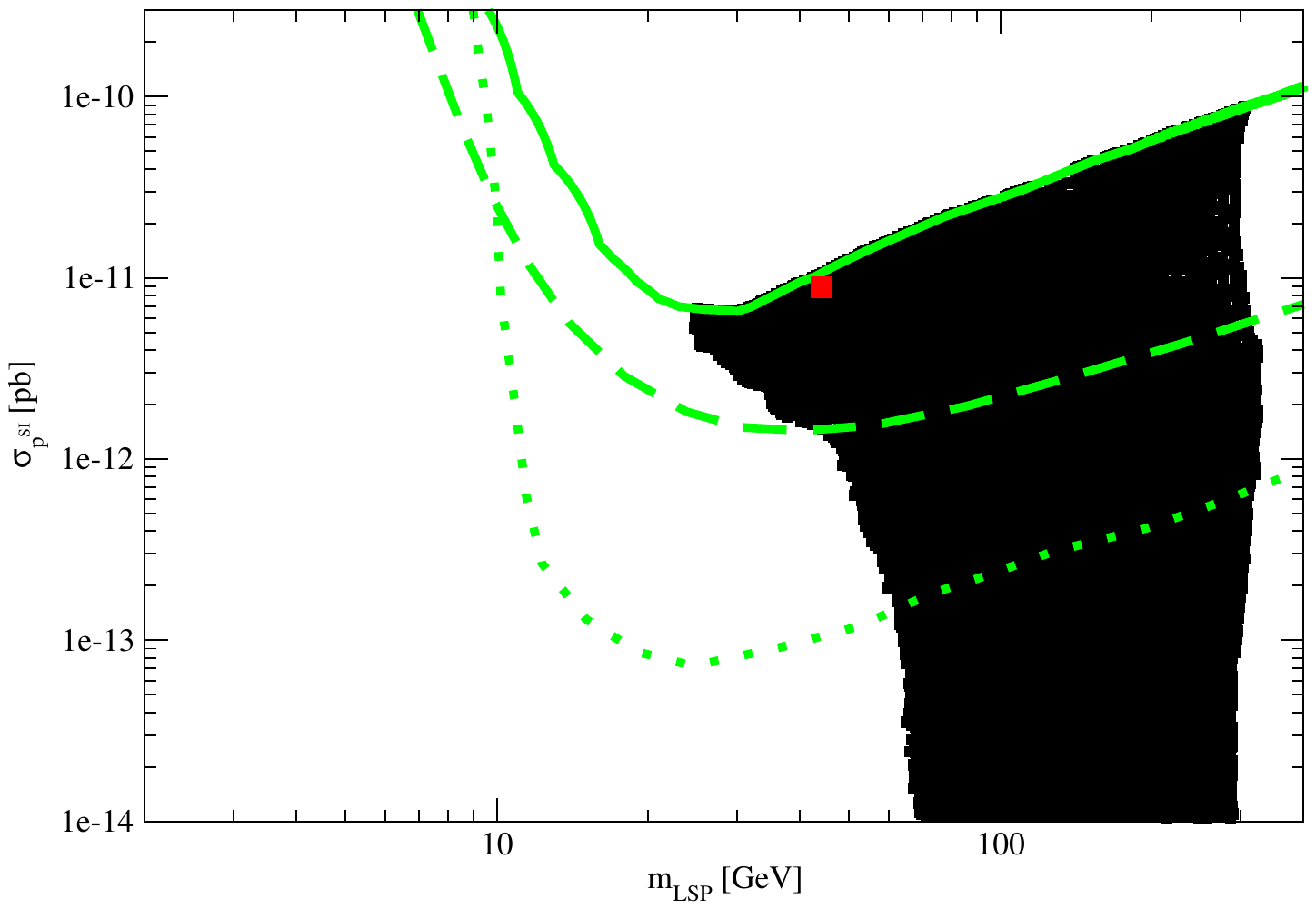}
\\
\includegraphics[scale=0.3]{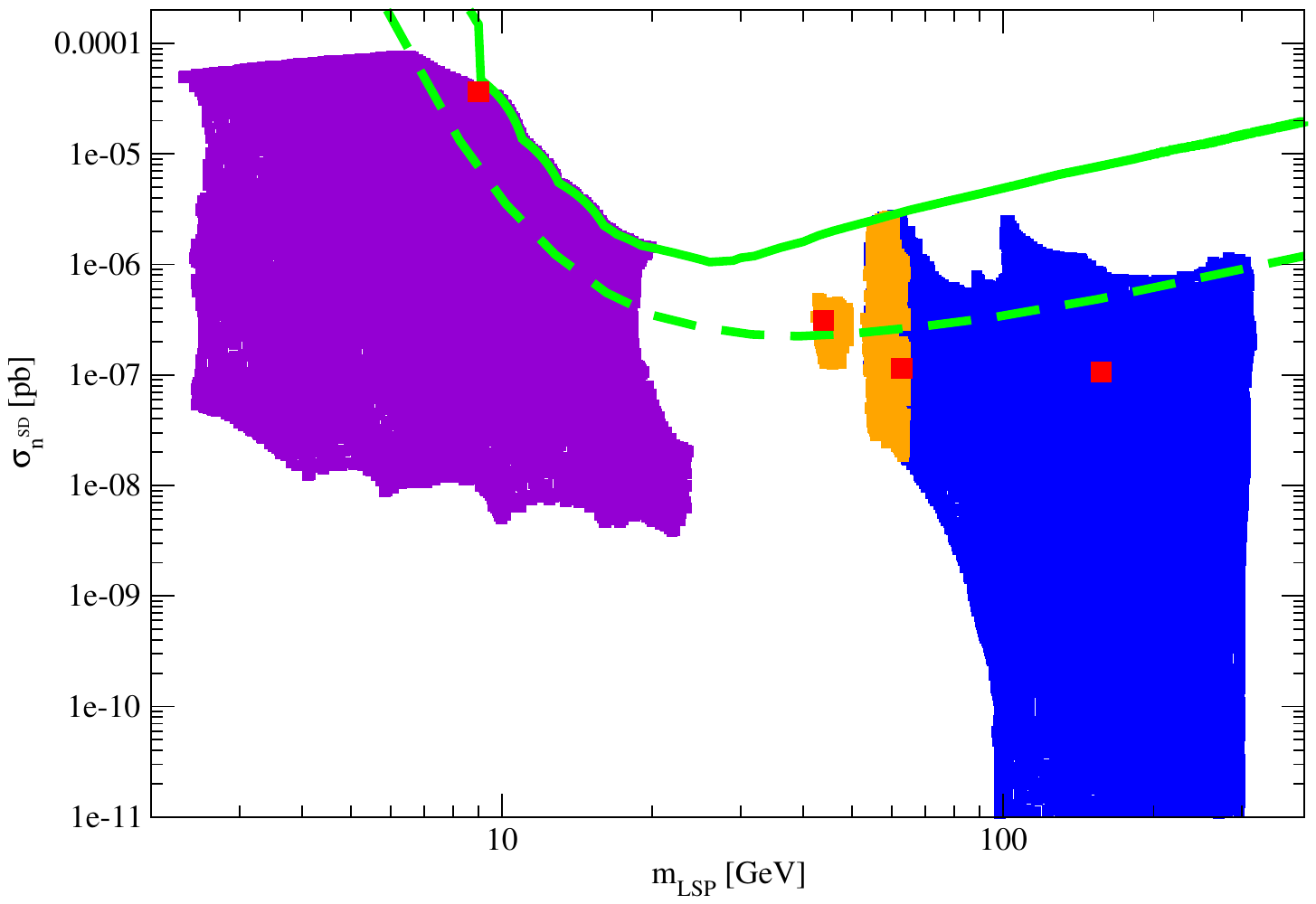}
   & 
\includegraphics[scale=0.3]{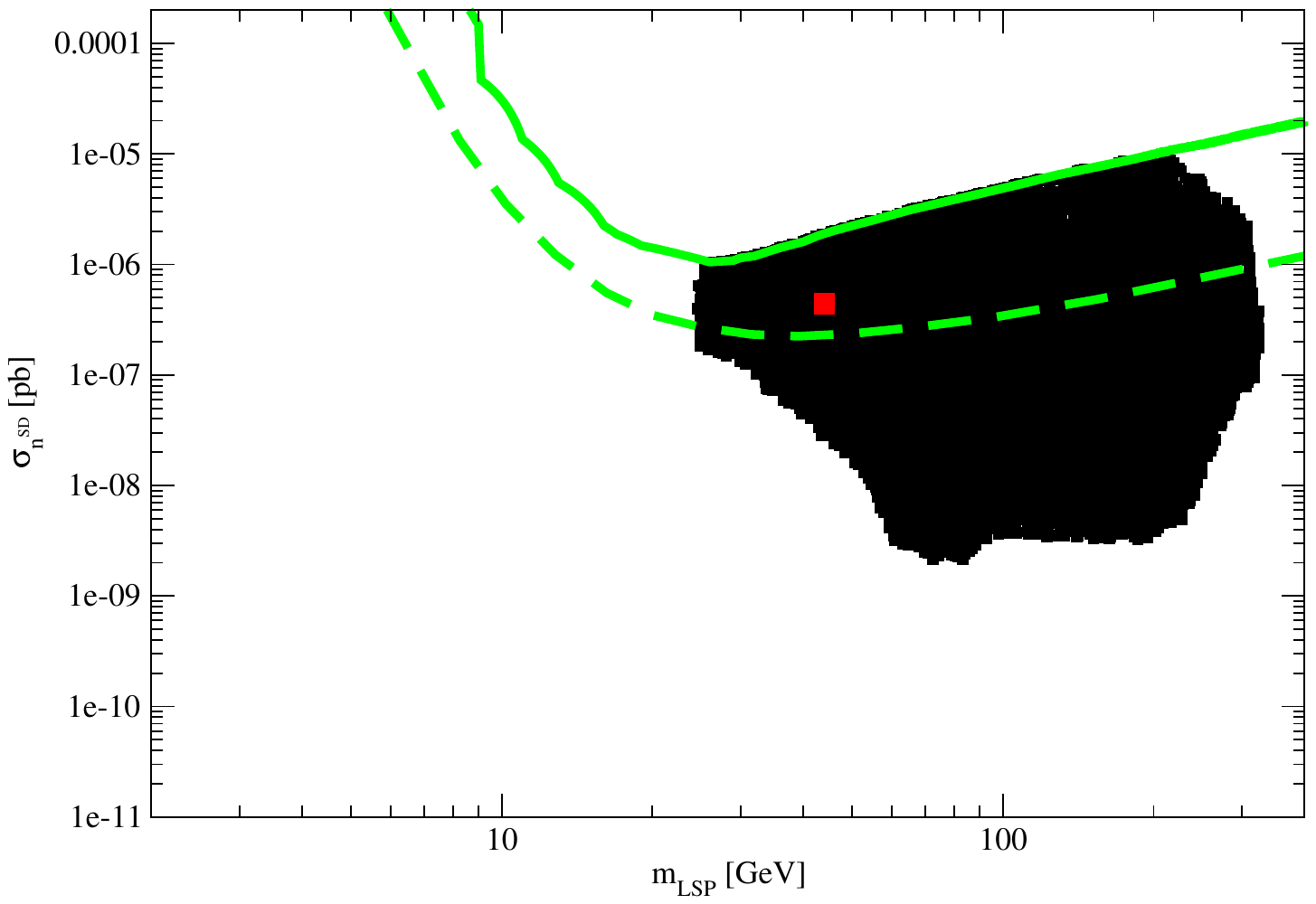}
\end{tabular}
\end{center}
\caption{$\sigma_{p}^{SI}$ and $\sigma_{n}^{SD}$ for points in the NMSSM with the correct relic density and satisfying all constraints. The colors denote the dark matter annihilation mechanisms as in Fig.~1. The full green line indicates the present upper limits from LZ \cite{LZ:2022lsv}, the dashed green line the future prospects on upper limits from LZ from \cite{LZ:2018qzl}, the dotted green line the neutrino floor from \cite{Billard:2013qya}.}
\label{fig:4}
\end{figure}

In Fig.~5 we show the mass $M_{A1}$ of the mostly singlet-like pseudoscalar, which is $\sim 2\times M_{LSP}$ for points in the $A_1$-funnel. For $M_{A1} \lsim 60$~GeV, $A_1$ may be visible  in decays of $H_{SM}$, see the latest study by CMS in \cite{CMS:2024uru}. Accordingly $A_1$ may be visible (but not necessarily) for all dark matter annihilation mechanisms.

\begin{figure}[ht!]
\begin{center}
\hspace*{-5mm}
\begin{tabular}{cc}
\includegraphics[scale=0.3]{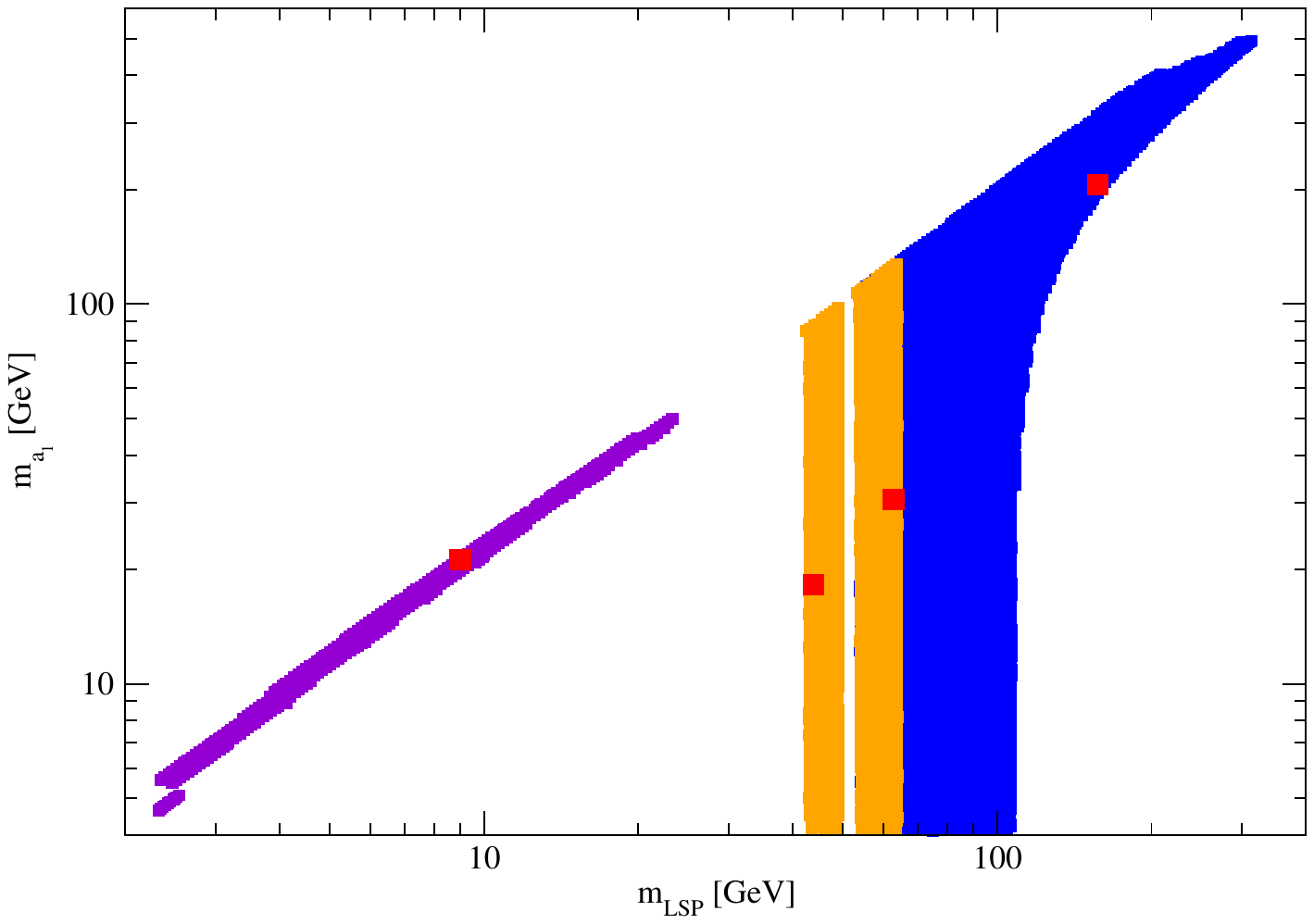}
   & 
\includegraphics[scale=0.3]{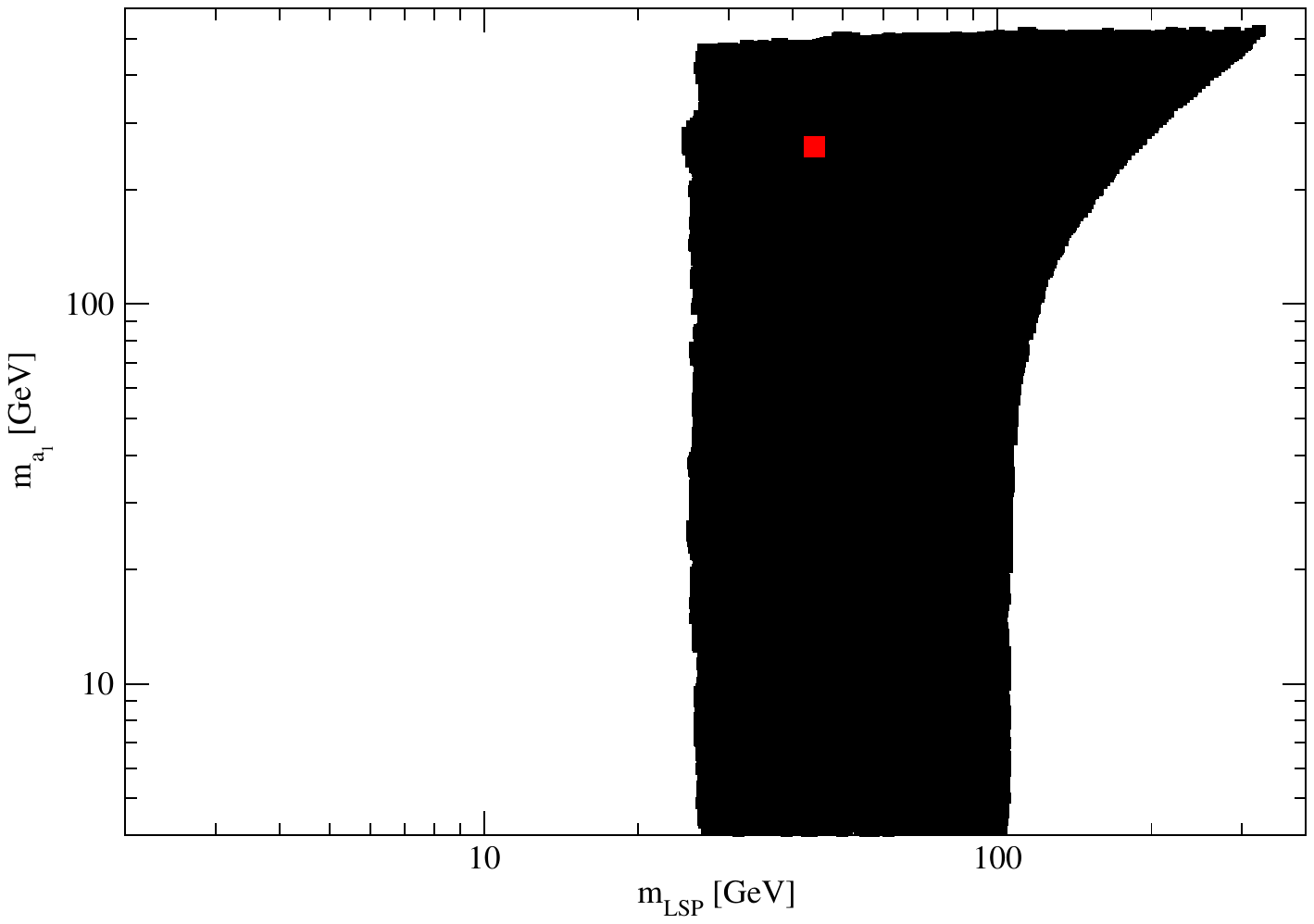}
\end{tabular}
\end{center}
\caption{$M_{A1}$ for points in the NMSSM satisfying all constraints. The colors denote the dark matter annihilation mechanisms as in Fig.~1.}
\label{fig:5}
\end{figure}


In Tables 2-4 we show the properties of some benchmark points (BPs) with distinct dark matter annihilation processes: Input parameters in Table~2, some Higgs masses and the nature of light electroweakinos in Table~3, and the properties shown in Figs.~1-5 in Table~4.

BP1 corresponds to dark matter annihilation via the $A_1$ funnel, accordingly the mass of the singlino-like LSP is close to $M_{A1}/2$. BP2 and BP3 correspond to dark matter annihilation via the $H_{95}$ and $H_{SM}$ funnels, respectively. Accordingly the masses of the singlino-like LSP are close to $M_{H_{95}}/2$ and $M_{H_{SM}}/2$, respectively. BP4 corresponds to co-annihilation with a slepton  and/or the second-lightest bino-like neutralino; in fact several possibilities are included here. $\sigma_{p}^{SI}$ for BP4 is below the neutrino floor. Finally BP5 corresponds to a bino-like LSP which is possible in the NMSSM as well as in the~MSSM.

\begin{table} [ht!]
\begin{center}
\begin{tabular}{| c | c | c | c | c | c |}
\hline
  & BP1  & BP2 & BP3 & BP4  & BP5  \\
\hline
$\lambda$       &$0.458$ & $0.242$ & $0.205$ & $0.0223$ & $0.0590$ \\
\hline
$\kappa$        &$-5.74\times 10^{-3}$ &$-7.26\times 10^{-3}$&$-8.06\times 10^{-3}$ & $9.63\times 10^{-3}$ & $7.75\times 10^{-3}$ \\
\hline
$A_\lambda$     &$3970$ &$4999$ & $4544$ & $-3390$& $-4995$ \\
\hline
$A_\kappa$      &$34.9$ &$11.0$& $11.0$& $183$ & $252$ \\
\hline
$\mu_{\rm eff}$ & $404$ & $736$ & $794$& $-179$ & $-681$\\
\hline
$\tan\beta$     & $9.76$ & $6.25$ & $5.13$ & $7.01$ & $8.94$  \\
\hline
$M_1$           & $163$ & $2065$ & $76.6$ & $196$ & $45.2$ \\
\hline
$M_2$           & $111$ & $100$ & $96.7$ & $-502$ & $-106$\\
\hline
$M_3$           & $2735$ & $4429$ & $1201$ & $3505$ & $825$\\
\hline
\end{tabular}
\caption{NMSSM specific input parameters and bino ($M_1$), wino ($M_2$) and gluino ($M_3$) masses for the five benchmark points. All dimensionful parameters are given in GeV.}
\end{center}
\label{tab:2}
\end{table}

\begin{table}[ht!]
\begin{center}
\begin{tabular}{| c | c | c | c | c | c |}
\hline
          & BP1  & BP2 & BP3 & BP4  & BP5  \\
\hline
$M_{H3}$ &$3966$ & $4852$ & $4367$ & $2069$ & $5592$ \\
\hline
$M_{A1}$ &$21$ & $18$ & $31$ & $206$ & $260$\\
\hline
LSP      & singl. & singl. & singl. & singl. & bino \\
\hline
$M_{\text {LSP}}$ & $9.0$ & $43.9$ & $62.9$& $157$ & $44.1$\\
\hline
NLSP     & wino$^\pm$ & wino$^\pm$ & bino & bino & wino$^0$  \\
\hline
$M_{\text {NLSP}}$  & $115$ & $107$ & $74.8$ & $164$ & $113$ \\
\hline
Slepton  & $\tilde{\nu}_\tau$ & $\tilde{\nu}_e$ & $\tilde{e}_{R.h.}$ & $\tilde{\nu}_e$ & $\tilde{\nu}_\tau$  \\
\hline
$M_{\text{Slepton}}$ & $140$ & $171$ & $255$ & $165$ & $93.4$ \\
\hline
\end{tabular}
\caption{Some BSM Higgs masses, the nature of the LSP, NLSP, the lightest slepton, and the corresponding masses for five BMpoints. All dimensionful parameters are given in GeV.}
\end{center}
\label{tab:3}
\end{table}


\begin{table}[ht!]
\begin{center}
\begin{tabular}{| c | c | c | c | c | c |}
\hline
  & BP1  & BP2 & BP3   & BP4  & BP5\\
\hline
$\mu^{LEP}_{bb}$  & $1.18\times 10^{-2}$ & $0.115$ & $0.106$ & $0.105$ & $0.116$  \\

$\mu^{LHC}_{\gamma\gamma}$ & $0.331$ & $8.01\times 10^{-2}$ & $8.01\times 10^{-2}$ & $8.75\times 10^{-2}$ & $0.130$ \\

$\mu^{LHC}_{\tau\tau}$  & $1.28\times 10^{-2}$ & $0.112$ & $0.102$ & $9.34\times 10^{-2}$ & $0.117$  \\

$\sigma_{p}^{SI}$ & $3.43\times 10^{-11}$ & $4.98\times 10^{-12}$ & $1.59\times 10^{-12}$ & $1.49\times 10^{-13}$ & $8.90\times 10^{-12}$\\

$\sigma_{n}^{SD}$ & $3.66\times 10^{-5}$ & $3.12\times 10^{-7}$ & $1.15\times 10^{-7}$ & $1.06\times 10^{-7}$ & $4.40\times 10^{-7}$ \\

\hline
\end{tabular}
\caption{$\mu^{LEP}_{bb}$, $\mu^{LHC}_{\gamma\gamma}$, $\mu^{LHC}_{\tau\tau}$, $\sigma_{p}^{SI}$ and $\sigma_{n}^{SD}$ (the latter in pb) for the five benchmark points. $\sigma_{p}^{SI}$ for BP4 is below the neutrino floor.}
\end{center}
\label{tab:4}
\end{table}


\section{Conclusions}

The NMSSM allows to describe simultaneously three phenomena related to physics beyond the Standard Model: A dark matter relic density of the amount observed in cosmology, but compatible with the absence of direct detection signals; a deviation of the muon anomalous magnetic moment by 4.2$\,\sigma$ from its value in the Standard Model; anomalies which indicate the presence of an extra scalar of a mass near 95~GeV in the $bb$~channel at LEP, and in $\gamma\gamma$ and $\tau\tau$ channels at the LHC. The latter would be impossible to describe within the MSSM. The LSP may have a mass well below 100~GeV, and at least part of the electroweak sparticle sector (sleptons, winos, bino) is typically relatively light. Such sparticles are difficult to rule out at the LHC, and generate complicated decay cascades also for the heavier strongly interacting sparticles. Their verification may then be a demanding task for the future.

\end{document}